\documentclass[sigconf]{acmart}

\usepackage{multirow}
\usepackage{algorithm}
\usepackage{algpseudocode}
\AtBeginDocument{%
  }

\setcopyright{acmlicensed}
\copyrightyear{2018}
\acmYear{2018}
\acmDOI{XXXXXXX.XXXXXXX}
\acmConference[Conference acronym 'XX]{Make sure to enter the correct
  conference title from your rights confirmation email}{June 03--05,
  2018}{Woodstock, NY}
\acmISBN{978-1-4503-XXXX-X/2018/06}

\begin{document}

\author{Kuan-Lin Chu}
\affiliation{%
  \institution{CITI, Academia Sinica}
  \country{Taiwan, ROC}
}

\author{Jun-Cheng Chen}
\affiliation{%
  \institution{CITI, Academia Sinica}
  \country{Taiwan, ROC}
}

\author{Chun-Shien Lu}
\affiliation{%
  \institution{IIS, Academia Sinica}
  \country{Taiwan, ROC}
}

\title{SSTMark: Robust Training-Free Semantic-Level\\ Speech Watermarking}



\begin{abstract}

As speech generation models become increasingly realistic and widely accessible, concerns about the misuse, attribution, and governance of synthetic speech continue to grow. Watermarking provides a practical way to make synthesized speech traceable and verifiable. Most existing speech watermarking methods embed watermark information into signal-level representations, such as waveforms or spectrograms. 
Under sufficiently strong distortions, the embedded watermark may be weakened or destroyed, leading to degraded detectability. In this paper, we propose SSTMark, a training-free speech watermarking framework that operates at the semantic level through text watermarking.
Unlike conventional signal-level watermarking methods, SSTMark encodes watermark information into the semantic content conveyed by generated speech, and detects the watermark from the recovered linguistic content. Experiments on AudioMarkBench demonstrate that SSTMark exhibits the strongest average robustness. Compared with the state-of-the-art baselines at a fixed false positive rate of 1\%, SSTMark improves the average detection rate by 4.6\% and 16.9\% on signal-processing edits and compression edits, respectively.

\end{abstract}




\maketitle

\begin{figure}[!htbp]
  \centering
  \includegraphics[width=0.4\textwidth]{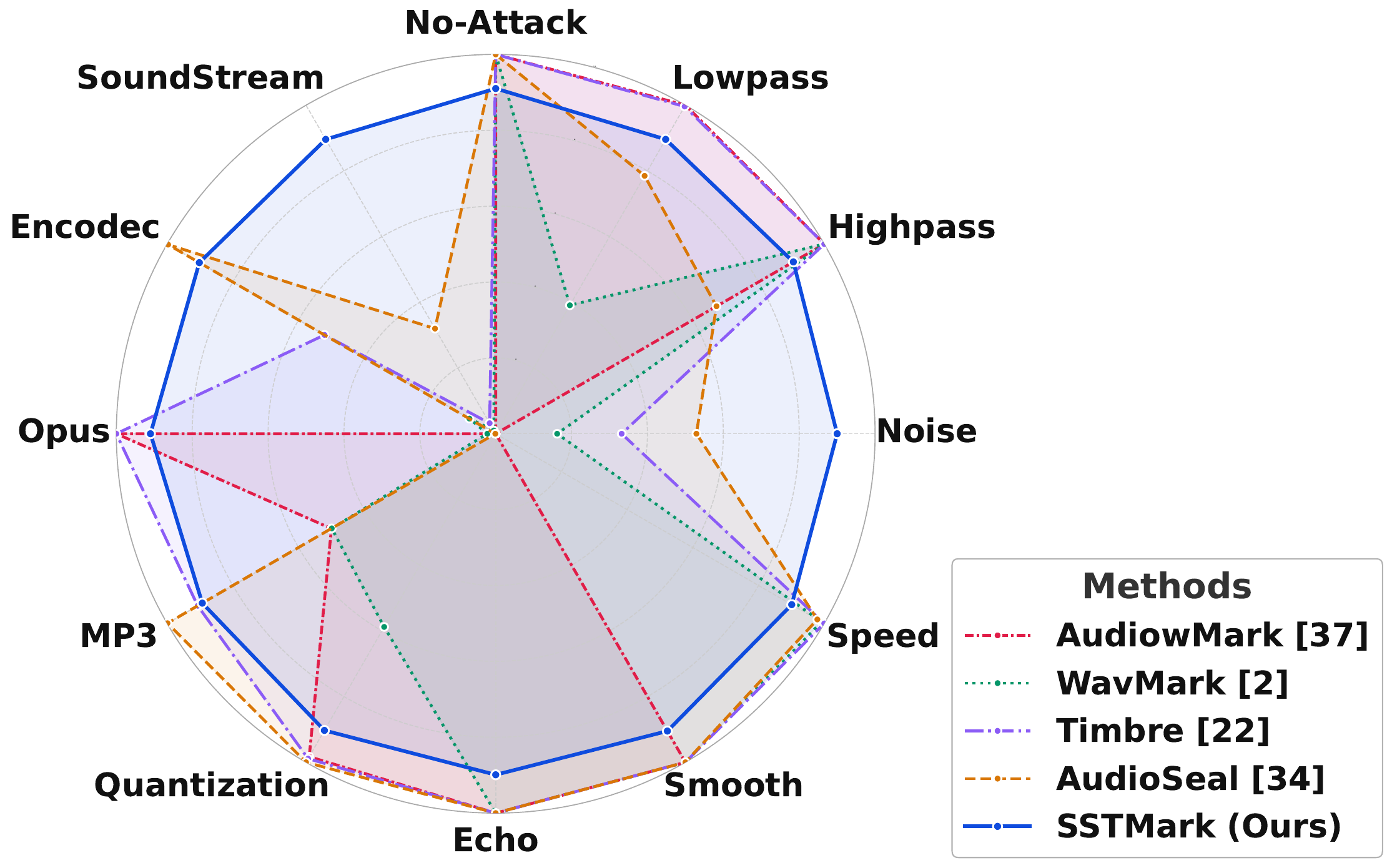}
  \caption{Robustness comparison under AudioMarkBench attacks~\cite{liu2024audiomarkbench} at a fixed false positive rate (FPR) of 1\%. Our method employs semantic-level watermark manipulation through speech-to-text (STT) and text-to-speech (TTS), which contributes to its strong average robustness on AudioMarkBench. Each vertex of the radar chart corresponds to one attack type, and the score on each axis denotes the average detection rate across different attack strengths.}
  \label{Fig:radar}
\end{figure}

\section{Introduction}
\label{sec:introduction}

\subsection{Background}
Recent advances in speech generative models~\cite{zhang2023speechgpt, chen2024vall, du2024cosyvoice, nguyen2025spirit, wang2025vocalnet, Chen2025neural} have made synthetic speech increasingly natural, expressive, and speaker-realistic, enabling a wide range of downstream applications. However, these advances have also made it more difficult to reliably distinguish AI-generated audio from authentic speech, raising serious risks of misuse, including impersonation, scams, misinformation dissemination, and political manipulation~\cite{li2024audio}.
%
%
Recent reports of AI-based impersonation, voice-cloning scams, and rising financial losses from deepfake-related fraud~\cite{guardian_rubio_ai_2025,fox13_ai_daughter_2025,esecurityplanet_deepfake_2025} further highlight the urgent need for reliable provenance mechanisms for AI-generated speech.

A major approach to mitigating these risks is speech deepfake detection~\cite{li2024audio, guo2024audio, jung2022aasist, liu2023asvspoof}, which identifies synthetic speech from artifacts or distributional inconsistencies in the audio signal. However, this paradigm is fundamentally passive, since authenticity is assessed only after the content has been generated and released. Watermarking provides a proactive alternative by embedding verifiable provenance evidence into the generated content itself. Existing speech watermarking methods can be broadly categorized as post-hoc watermarking~\cite{san2024proactive, chen2023wavmark, liu2024detecting, liu2025xattnmark} or generative watermarking~\cite{liu2024groot}. In both cases, watermarking offers the possibility of equipping generated speech with machine-verifiable provenance cues.

Existing speech watermarking methods still face a major robustness challenge in realistic deployment. Most representative approaches embed and verify watermark signals in signal-level representations, such as waveforms or spectrograms~\cite{liu2024audiomarkbench, odeep, san2024proactive, chen2023wavmark, liu2024detecting, zong2025audiomarknet, liu2024groot}, so successful verification depends on preserving low-level acoustic patterns. In practice, however, generated speech is often subjected to additive noise, filtering, resampling, compression, speed perturbation, neural codec reconstruction, and other post-processing operations~\cite{li2024audio, li2025survey, liu2024audiomarkbench}. Although such transformations may preserve intelligibility and the core message conveyed by the speech, they can substantially alter acoustic details and thereby reduce or even remove the watermark evidence used for verification.

\subsection{Motivation}

From the perspective of a malicious user, generated speech intended for deceptive use is often further manipulated before release, rather than being used in its original form~\cite{li2024audio, li2025survey, liu2024audiomarkbench}. Although these operations can substantially alter acoustic details, the spoken message must typically remain intelligible; otherwise, the manipulated speech would fail to convey the intended deceptive content. Consequently, low-level acoustic details may be heavily perturbed, while the semantic content often remains relatively stable.

This observation suggests that generated speech is not only an acoustic signal, but also a carrier of semantic content. Therefore, provenance need not be tied solely to acoustic form, but may also be associated with the linguistic realization of the speech itself. If a watermark can be bound to the semantic content conveyed by speech, then verification may rely less on the exact preservation of acoustic details and more on whether sufficient textual content can still be recovered from the speech after distortion. This motivates us to rethink speech watermarking not only as an acoustic-level signal hiding problem, but also as a semantic-level provenance problem.

\subsection{Overview and Contributions}

To address above challenges, we propose a \textbf{S}emantic-level \textbf{S}peech watermarking framework built
upon \textbf{T}ext Water\textbf{Mark}ing (SSTMark), which binds the watermark to the linguistic content conveyed by generated speech. Given a prompt-conditioned spoken response, SSTMark first recovers its textual content through speech transcription, injects watermark through semantically constrained text rewriting, and converts the rewritten text back into speech via text-to-speech resynthesis. During verification, the received speech is transcribed again and checked for watermark evidence from the recovered text. In this way, the watermark is not carried primarily by signal-level patterns, but by controlled textual realizations that are expected to remain recoverable as long as the speech content is still intelligible. Because SSTMark intentionally rewrites transcripts and resynthesizes speech, its utility should be evaluated not only in terms of robustness, but also in terms of semantic preservation, speaker consistency, and audio quality.
SSTMark achieves more consistent robustness across benchmark perturbations~\cite{liu2024audiomarkbench}, as illustrated in Fig. \ref{Fig:radar}. 

Our contributions are summarized as follows:

\begin{itemize}
    \item We propose a training-free semantic-level speech watermarking framework that leverages a speech--text conversion architecture to inject watermark evidence into the semantic content conveyed by speech.

    \item Different from prior studies, the adopted fidelity evaluation protocol is tailored to semantic-level speech watermarking. Specifically, rather than emphasizing waveform-level similarity, we analyze how transcript-level rewriting affects textual preservation, speaker consistency, and speech quality after speech resynthesis.


    \item We perform robustness evaluations under several benchmark perturbations. Experimental results show that our framework achieves uniform robustness under benchmark settings and retains a certain degree of detection capability under several severe attacks, highlighting the potential of semantic-level watermarking for more resilient provenance verification of generated speech.
\end{itemize}

\section{Related Work}
\label{sec:related_work}

\subsection{Speech Watermarking}
Recent speech watermarking research has explored deep learning for improving robustness, detectability, and localization. Many of them perform both watermark embedding and verification in the acoustic domain, while more recent studies have begun to explore generation-stage or intermediate-representation watermarking.

Among signal-level methods, including AudiowMark~\cite{westerfeld2020audiowmark}, WavMark~\cite{chen2023wavmark}, Timbre~\cite{liu2024detecting}, AudioSeal~\cite{san2024proactive}, XAttnMark~\cite{liu2025xattnmark}, DiscreteWM \cite{ji2025speech}, and AudioMarkNet~\cite{zong2025audiomarknet}, they mainly differ in the representations and architectures used for watermarking.
Some recent methods explore watermarking through generation-time mechanisms. GROOT~\cite{liu2024groot} performs in-generation watermarking by integrating watermark encoding into the diffusion-based audio synthesis process. 
ALIGNED-IS~\cite{wurobust, wu2025watermark}, a watermark for autoregressive audio generation models that addresses retokenization mismatch by clustering acoustically similar tokens.
AudioMarkBench~\cite{liu2024audiomarkbench} provides a systematic benchmark for evaluating the robustness of existing audio watermarking methods under watermark-removal and watermark-forgery attacks.

Overall, existing speech watermarking methods have primarily focused on acoustic, latent, or generation-stage representations. In contrast, much less attention has been paid to associating watermark evidence with the linguistic content conveyed by speech, which is the perspective adopted in this work.

\subsection{Text Watermarking}

Following the survey~\cite{10.1145/3691626}, text watermarking methods can be broadly divided into watermarking for existing text and watermarking for LLMs. In this work, we focus on research on watermarking for existing text, where already generated text is modified to produce a watermarked version.

For existing text, one line of work performs lexical-level rewriting under semantic constraints. Yang et al.~\cite{yang2023watermarking} replace selected words with context-aware synonyms and detect watermark presence through a statistical bias toward bit-1 words. Yoo et al.~\cite{yoo2023robust} extend this setting to multi-bit watermarking by selecting invariant positions and rewriting masked tokens with an infill model. Hao et al.~\cite{hao2025post} further adopt a post-hoc lexical substitution strategy that selects robust anchor positions from semantic and syntactic features and replaces words with paraphrase-based synonyms. Another line of work uses model-based regeneration or paraphrasing, as in WATERFALL~\cite{lau2024waterfall}, Xu et al.~\cite{xu2025robust}, and REMARK-LLM~\cite{zhang2024remark}, which rewrite the text more globally rather than through only local word substitution.


Overall, prior text watermarking studies show that both linguistic rewriting and generation-time control can embed detectable watermark evidence in text. However, they typically treat text as the final output modality, whereas our work uses text watermarking as an intermediate mechanism for provenance verification in generated speech.

\section{Preliminaries}
\label{sec:preliminaries}
In this section, we briefly introduce the auxiliary modules used in our framework.

\subsection{Speech-Text Conversion Models}
\label{sec:prelim_speech_text}
The proposed framework relies on two auxiliary modules to connect the speech and text domains: a speech-to-text (STT) model and a text-to-speech (TTS) model. The STT model, denoted by $\mathbf{A}(\cdot)$, converts a speech signal into its transcript through automatic speech recognition~\cite{radford2023robust, hsu2021hubert, gulati2020conformer}. In practice, the linguistic content of generated speech may not always be explicitly exposed in textual form during generation. The STT model therefore provides a practical mechanism for recovering a manipulable textual representation of the semantic content conveyed by speech.

Conversely, the TTS model, denoted by $\mathbf{T}(\cdot)$, converts a watermarked transcript back into speech~\cite{du2024cosyvoice, chen2025f5, casanova2024xtts, Chen2025neural}. This step is necessary because the final output of our system remains a speech signal rather than text. Although watermarking is carried out in the text space, the TTS model is needed to map the resulting watermarked transcript back into the speech domain for downstream use.

\subsection{Text Watermarking Module}
\label{sec:prelim_text_wm}
In our framework, speech is first converted into text for watermarking. Since text watermarking plays only an intermediate role in our method, we adopt the post-processing approach of Yang et al.~\cite{yang2023watermarking}, which injects watermark information by rewriting existing text under semantic constraints. 
Let $x=(x_1,x_2,\dots,x_n)$ denote an input text sequence, let $x_{\mathrm{wm}}$ be the resulting watermarked text, and let $z$ denote the z-score produced by the detector to quantify the statistical evidence of watermark presence in the text.

Specifically, each token $x_i$ is assigned a binary encoding $b_i$, which depends on the current token and preceding token, so that the same token may correspond to bit-0 or bit-1 in different contexts:
\begin{equation}
\label{eq:bit_encoding}
b_i=\mathrm{RandomBinary}(h(x_i)\oplus h(x_{i-1})), \qquad 2\leq i\leq n,
\end{equation}
where $\oplus$ denotes bitwise XOR, $h(\cdot)$ denotes a string hash function, $b_i\in\{0,1\}$, and $\mathrm{RandomBinary}(\cdot)$ maps the resulting hash value to a binary value. Owing to the near-uniform behavior induced by this hash-based randomization, the binary encodings of non-watermarked text are expected to be approximately balanced between bit-0 and bit-1.

Moreover, if $x_i$ passes the part-of-speech filter and its binary encoding is bit-0, synonym candidates are generated for possible replacement. 
In~\cite{yang2023watermarking}, a BERT-based context-aware procedure is used to generate synonym candidates, which are filtered using both sentence-level and word-level semantic constraints. 
Let $\tau_{\mathrm{sent}}$ and $\tau_{\mathrm{word}}$ denote the sentence-level and word-level similarity thresholds, respectively. 
For each token $x_i$,~\cite{yang2023watermarking} uses a BERT-based context-aware procedure to generate an initial candidate set of synonym substitutions, denoted by $C=\{s_1,s_2,\dots,s_K\}$. Then, a so-called filtered candidate set is retained as:
\begin{equation}
\label{eq:candidate_filter}
C'=\{s \in C \mid S_{\mathrm{sent}}(s,x_i)\ge \tau_{\mathrm{sent}} \;\text{and}\; S_{\mathrm{word}}(s,x_i)\ge \tau_{\mathrm{word}}\},
\end{equation}
where $S_{\mathrm{sent}}(s,x_i)$ and $S_{\mathrm{word}}(s,x_i)$ denote the sentence-level and word-level similarity scores, respectively. 

Finally, binary encoding is recomputed only for each candidate in the filtered candidate set $C'$, using the same rule as Eq.~(\ref{eq:bit_encoding}). 
Among the candidates whose binary encoding is bit-1, the one with the highest word-level semantic similarity to the original token $x_i$ is selected. Replacing $x_i$ with the selected candidate increases the proportion of bit-1 encodings in the rewritten text. By repeating this process, a statistical bias is injected into the rewritten text, yielding the final watermarked text $x_{\mathrm{wm}}$. 
We denote this sequential rewriting procedure by the abstract watermarking operator $\mathbf{W}(\cdot)$.

For detection, the corresponding detector examines whether the observed binary encodings deviate significantly from the approximately balanced distribution. Let $n_1$ denote the number of tokens encoded as bit-1 and let $N$ denote the total number of encoded tokens. Equivalently, let $\hat{p}=n_1/N$ be the observed proportion of bit-1 encodings. Thus, watermark detection is formulated as a hypothesis test with null hypothesis $H_0$ that the observed binary encodings follow the approximately balanced distribution expected from non-watermarked text. The detector computes the test statistic
\begin{equation}
\label{eq:zscore}
z=\frac{\hat{p}-p_0}{\sqrt{p_0(1-p_0)/N}},
\end{equation}
where $p_0=0.5$ is the expected proportion of bit-1 under the null hypothesis. A larger $z$-score indicates stronger watermark evidence, and watermark presence is determined by thresholding this score. We denote this score-based verification procedure by the abstract detector operator $\mathbf{D}(\cdot)$.

\begin{figure*}[!htbp]
  \centering
  \includegraphics[width=0.8\textwidth]{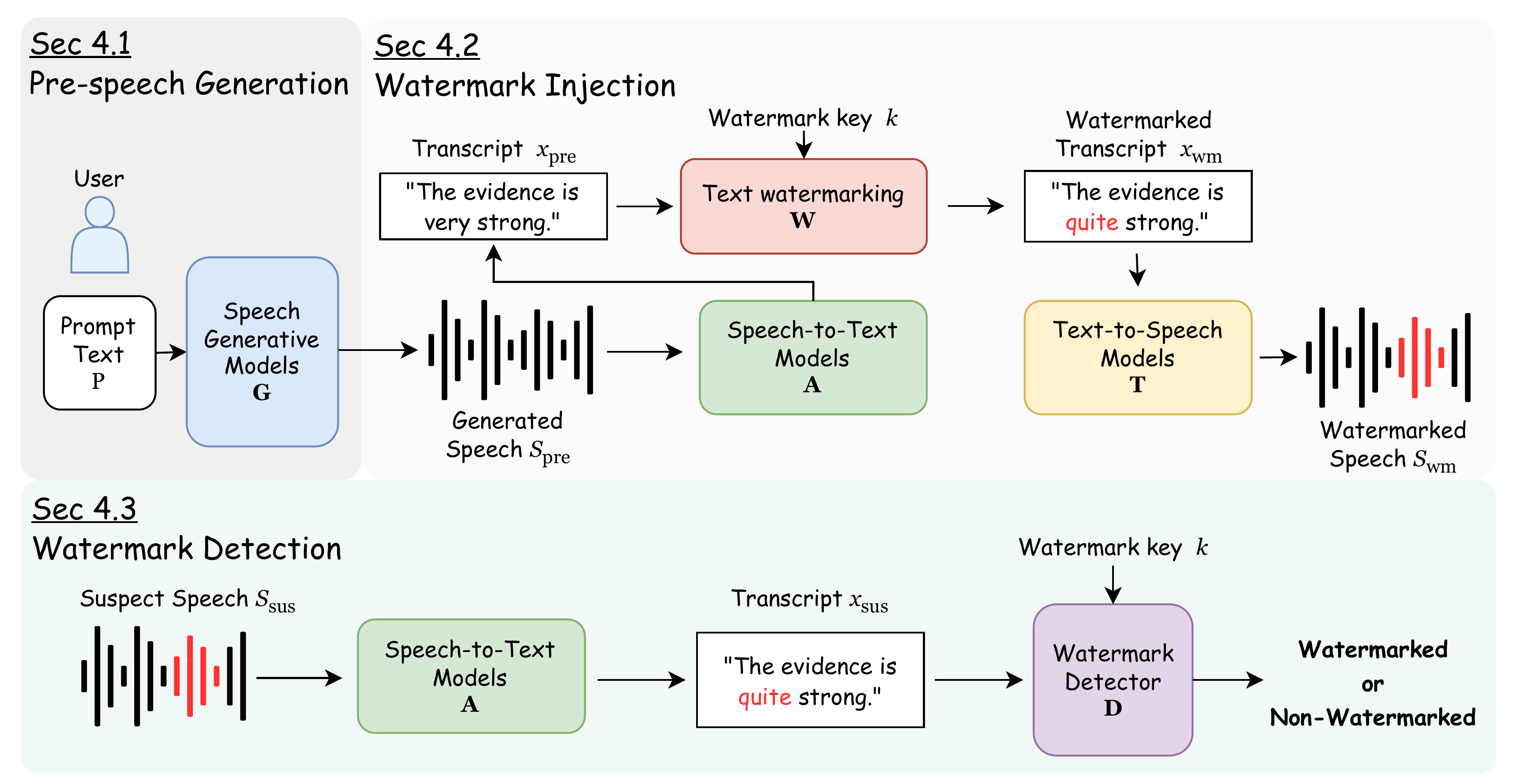}
  \caption{Overview of SSTMark. Given a prompt text $p$, the speech generative model $\mathbf{G}$ first generates pre-speech $s_{\mathrm{pre}}$, which is transcribed by the speech-to-text model $\mathbf{A}$ into transcript $x_{\mathrm{pre}}$. The text watermarking module $\mathbf{W}$, conditioned on watermark key $k$, rewrites $x_{\mathrm{pre}}$ into watermarked transcript $x_{\mathrm{wm}}$, which is then converted by the text-to-speech model $\mathbf{T}$ into watermarked speech $s_{\mathrm{wm}}$. For detection, a suspect speech sample $s_{\mathrm{sus}}$ is transcribed by $\mathbf{A}$ into transcript $x_{\mathrm{sus}}$, and the watermark detector $\mathbf{D}$ verifies whether watermark information associated with $k$ is present.}
  
  \label{Fig:Framework}
\end{figure*} 

\section{Proposed Method: SSTMark}
\label{sec:proposed_method}
SSTMark is designed to associate watermark information with the semantic content conveyed by generated speech. The overall watermarking pipeline consists of three stages: pre-speech generation, watermark injection, and watermark detection, as illustrated in Fig.~\ref{Fig:Framework}. We first provide an overview of the proposed method, and then present the details of each stage in the following subsections.

Our framework is a post-hoc speech watermarking approach: Rather than modifying the internal synthesis process of the speech generator, it first obtains a spoken response from the generative model and then applies watermark injection to the recovered response content. Accordingly, SSTMark differs from conventional speech watermarking methods in two main aspects. First, our method is training-free and does not require attack-aware optimization during watermark embedding. In contrast, some prior speech watermarking methods enhance robustness by incorporating noise layers or simulated attack modules during training, primarily to improve performance under known signal distortions.

Second, our method embeds and verifies watermark information at the semantic level rather than the signal level. Since a range of common audio distortions primarily affect acoustic details, our design aims to bind watermark information to the semantic content carried by speech, thereby reducing reliance on the exact preservation of acoustic features. Conceptually, our method reformulates speech watermarking as a semantic-level hiding problem by performing watermark injection and detection on the textual content recovered from speech. As a result, watermark detection no longer depends on recovering fine-grained acoustic perturbations, but rather on recovering sufficient textual evidence from the speech.

To improve the practical usability of SSTMark, we further introduce a watermark key used during embedding and detection. The role of this key is not to encrypt the speech signal itself, but to control how watermark patterns are instantiated in the text space. Under this design, the same speech content may induce different token-level watermark patterns under different keys, making the detection process dependent on the shared key.

The full algorithmic procedure of our method is provided in Appendix~\ref{sec:pseudocode}.

\subsection{Pre-speech Generation}

Given an input prompt $p$, a speech generative model $\mathbf{G}_{\mathrm{sp}}(\cdot)$ first generates a response speech sample, referred to as pre-speech $s_{\mathrm{pre}}$:
\begin{equation}
\label{eq:pre-speech}
s_{\mathrm{pre}} = \mathbf{G}_{\mathrm{sp}}(p).
\end{equation}
This pre-speech represents the initial spoken response produced by the generative model and serves as the starting point of the proposed watermarking pipeline.

The role of this stage is to generate an intermediate speech sample before watermark injection. Since the proposed method binds watermark information to semantic content, it requires access to a textual representation of the content conveyed by the generated speech. However, in practical speech generation, such information may not always be explicitly exposed in a directly manipulable text form. Therefore, rather than assuming direct access to an internal textual response, we first obtain the model output in speech form and then recover its transcript in the subsequent stage through automatic speech recognition.

Under this design, the pre-speech $s_{\mathrm{pre}}$ serves as the source signal for downstream transcription, which subsequently enables watermark injection in text space. More broadly, this stage allows the proposed framework to be applied as a post-hoc wrapper around existing speech generators. In particular, SSTMark does not require access to internal logits, hidden states, or decoder-side textual outputs of the speech generator. Instead, it only assumes that the generator can produce a speech waveform that can be transcribed by an external automatic speech recognition model. As a result, SSTMark is not restricted to a specific generator architecture, such as end-to-end speech language models, and can also be applied to other speech generation systems that output speech directly.

\subsection{Watermark Injection}

After pre-speech generation, the objective of watermark injection is to recover an explicitly manipulable text-space representation conveyed by speech, inject watermark information into the text space, and then map the watermarked transcript back into the speech domain. Accordingly, this stage consists of three steps: transcript extraction, text watermark injection, and speech resynthesis.

First, to recover the semantic content conveyed by the generated pre-speech, we apply an automatic speech recognition model $\mathbf{A}(\cdot)$ to pre-speech $s_{\mathrm{pre}}$:
\begin{equation}
\label{eq:STT}
x_{\mathrm{pre}} = \mathbf{A}(s_{\mathrm{pre}}),
\end{equation}
where $x_{\mathrm{pre}}$ denotes the transcript and is treated as the text-space representation conveyed by the pre-speech.

After obtaining the transcript $x_{\mathrm{pre}}$, we inject watermark information in the text space using the text watermarking operator introduced in Sec.~\ref{sec:prelim_text_wm}. Specifically, we follow the binary encoding rule in Eq.~\eqref{eq:bit_encoding} and later extend it to a key-dependent formulation by incorporating a watermark key $k$. In the text watermarking method described in Sec.~\ref{sec:prelim_text_wm}, the random bit is generated from the string hash value in Eq.~\eqref{eq:bit_encoding}. In our framework, we further incorporate the watermark key into this hash value construction, so that the resulting token-level watermark pattern is conditioned on both local textual context and the shared key used in embedding and detection. As a result, the same transcript may induce different token-level binary patterns under different watermark keys, while the embedding and detection procedures remain aligned when instantiated with the same key.

Concretely, by augmenting Eq.~(\ref{eq:bit_encoding}) with the key term $h(k)$, the binary encoding of token $x_i$ is defined as
\begin{equation}
\label{eq:keyed_token_bit}
b_i=\mathrm{RandomBinary}(h(k)\oplus h(x_i)\oplus h(x_{i-1})).
\end{equation}
Under the same modification, the binary encoding assigned to each candidate synonym during the selection process is also extended to the keyed setting.
Specifically, for a candidate synonym $s_j$, its binary encoding under the current context becomes
\begin{equation}
\label{eq:keyed_synonym_bit}
b_j=\mathrm{RandomBinary}(h(k)\oplus h(s_j)\oplus h(x_{i-1})).
\end{equation}
Thus, the bit assignment depends on both the local textual context and watermark key, rather than on the token identity alone. As a result, the same transcript may induce different token-level binary patterns under different watermark keys, while the detector can reconstruct the same key-conditioned bit-assignment rule when the same key is used.

Under this keyed binary encoding, the synonym filtering and replacement procedure described in Sec.~\ref{sec:prelim_text_wm} becomes a key-conditioned rewriting process, which we denote by $\mathbf{W}(\cdot;k)$. The resulting watermarked transcript $x_{\mathrm{wm}}$ is then derived from $x_{\mathrm{pre}}$ as:
\begin{equation}
\label{eq:watermark}
x_{\mathrm{wm}} = \mathbf{W}(x_{\mathrm{pre}}; k).
\end{equation}
In this step, watermark information is introduced into the text-space representation conveyed by $s_{\mathrm{pre}}$. The resulting $x_{\mathrm{wm}}$ is expected to preserve the semantic content of $x_{\mathrm{pre}}$ while carrying watermark-induced statistical bias that can be detected during verification.

Following the candidate filtering rule in Eq.~(\ref{eq:candidate_filter}), we adopt the original setting~\cite{yang2023watermarking} and fix the sentence-level similarity threshold at $\tau_{\mathrm{sent}}=0.83$, while using $\tau_{\mathrm{word}}$ to control watermark strength. A smaller $\tau_{\mathrm{word}}$ relaxes the word-level semantic constraint and allows more replacement candidates, thereby leading to stronger watermark injection, whereas a larger $\tau_{\mathrm{word}}$ imposes a stricter semantic constraint and results in weaker watermark injection.

Once the watermark has been injected into the transcript, the watermarked transcript $x_{\mathrm{wm}}$ is converted back into speech by a text-to-speech model $\mathbf{T}(\cdot)$, yielding the final watermarked speech:
\begin{equation}
\label{eq:TTS}
s_{\mathrm{wm}} = \mathbf{T}(x_{\mathrm{wm}}).
\end{equation}

\subsection{Watermark Detection}

The objective of watermark detection is to verify whether watermark information is present in a received speech sample by recovering its textual representation and performing watermark detection in the text space. Since our method binds watermark information to the semantic content conveyed by speech, watermark detection is conducted on the transcript.

Given a received speech sample $s_{\mathrm{sus}}$ that may be manipulated from $x_{\mathrm{wm}}$, we first apply the automatic speech recognition model $\mathbf{A}(\cdot)$ to obtain its transcript $x_{\mathrm{sus}}$ as:
\begin{equation}
\label{eq:STT_sus}
x_{\mathrm{sus}} = \mathbf{A}(s_{\mathrm{sus}}).
\end{equation}
It is worth noting that both the embedding-side transcript $x_{\mathrm{pre}}$ and detection-side transcript $x_{\mathrm{sus}}$ are obtained via external STT recovery, rather than directly observed as ground-truth textual responses. Accordingly, our framework does not require the transcript to exactly match  
the original response semantic content at the word level. Instead, it relies on the recovered transcripts preserving sufficient linguistic evidence such that the 
watermark-related bit-assignment bias introduced during rewriting remains consistently expressible under $\mathbf{W}(\cdot;k)$ and detectable by $\mathbf{D}(\cdot;k)$.

Because watermark embedding and detection share the same key-conditioned token encoding rule, the watermark evidence introduced during rewriting is manifested as a matched statistical bias in the recovered transcript and can therefore be re-evaluated by the detector under the same key. In this sense, $\mathbf{D}(\cdot;k)$ does not search for an arbitrary pattern in text, but specifically tests whether the transcript exhibits the same key-dependent bit-assignment bias induced by $\mathbf{W}(\cdot;k)$ during embedding.

After obtaining the transcript $x_{\mathrm{sus}}$, we apply the corresponding key-conditioned text watermark detector, denoted by $\mathbf{D}(\cdot; k)$, to evaluate whether watermark information associated with the key $k$ is present. Specifically, $\mathbf{D}(\cdot; k)$ follows the same hypothesis-testing formulation in Eq.~(\ref{eq:zscore}), except that the token encodings are generated using the key-conditioned rule adopted in this section, and thereby producing the watermark-related z-score
\begin{equation}
\label{eq:wm_detector}
z = \mathbf{D}(x_{\mathrm{sus}}; k).
\end{equation}
Here, $z$ is computed using the z-score formulation in Eq.~(\ref{eq:zscore}) under the same key-conditioned token encoding used during embedding.

\section{Experiments}
\label{sec:experiments}

\subsection{Experimental Setup}
\label{sec:experimental_setup}

\paragraph{Baseline methods.}
In the main experiments, we compare SSTMark with four state-of-the-art speech watermarking baselines, including AudiowMark~\cite{westerfeld2020audiowmark}, WavMark~\cite{chen2023wavmark}, Timbre~\cite{liu2024detecting}, and AudioSeal~\cite{san2024proactive}. XAttnMark~\cite{liu2025xattnmark} is not included in our comparison because no official code release is available before the paper submission deadline. 
For methods with publicly available code, we used the official pretrained models and default inference settings, and evaluated all methods on the same generated speech samples under identical attack conditions.

\paragraph{Speech generation and resynthesis.}
For pre-speech generation, we used SpeechGPT~\cite{zhang2023speechgpt} as the speech generative model $\mathbf{G}_{\mathrm{sp}}(\cdot)$ in Eq.~(\ref{eq:pre-speech}). Specifically, we used the complete LibriSpeech~\cite{panayotov2015librispeech} train-clean-100 split as the prompt source, and followed the speech generation setup described in the original SpeechGPT~\cite{zhang2023speechgpt}. 
Detailed generation instructions are provided in Appendix~\ref{sec:generation_template}. For speech transcription and resynthesis in SSTMark, we used Whisper Large-v3~\cite{radford2023robust} as the STT model $\mathbf{A}(\cdot)$ in Eq.~(\ref{eq:STT}) and CosyVoice~\cite{du2024cosyvoice} as the TTS model $\mathbf{T}(\cdot)$ in Eq.~(\ref{eq:TTS}). Whisper and CosyVoice were both used with their default decoding settings. For TTS synthesis, the reference audio was taken from the corresponding non-watermarked pre-speech sample. 
In our implementation, before feeding the rewritten text into the TTS model, we constrained input text length to at most 200 text tokens, resulting in an average synthesized speech duration of approximately 40 seconds. 
Before perturbation and detection, all audio samples were resampled to 16 kHz. All experiments were conducted on a server equipped with an NVIDIA Tesla V100-SXM2-32GB GPU.

\paragraph{Detection protocol.}
Because detection performance depends on the operating point, we evaluated all methods under matched false positive rate (FPR) constraints rather than using method-specific default thresholds. Accordingly, for SSTMark and the other baselines, detection thresholds were calibrated independently on clean non-watermarked speech samples to satisfy fixed FPR targets. Specifically, we used 1000 clean non-watermarked speech samples generated from prompts drawn from the LibriSpeech train-clean-100 split and selected thresholds to achieve FPRs of $1\times10^{-2}$ and $1\times10^{-3}$. We then reported the corresponding true positive rates (TPRs) at these two operating points. AudiowMark~\cite{westerfeld2020audiowmark} was treated separately, since its detection criterion is based on exact message matching, which leads to a relatively low false positive rate in practice. To reflect a realistic deployment scenario in which the detector does not know whether an incoming audio sample has been attacked or which distortion has been applied, thresholds calibrated on clean speech were kept fixed and directly applied to all attacked samples during robustness evaluation.

\paragraph{Fidelity metrics.}
To evaluate fidelity, we assessed whether the watermarked speech remains faithful to the original, non-watermarked output in terms of semantic content, speaker characteristics, and overall audio quality. Because the proposed method rewrites transcripts and resynthesizes speech, it may introduce substantial waveform variation and inconsistent waveform lengths even when semantic and perceptual properties are largely preserved. Accordingly, we do not emphasize waveform-level similarity and instead focus on semantic preservation and speaker consistency as more practical indicators of fidelity. We evaluated fidelity in both text and audio spaces. In the text space, we used BERTScore~\cite{zhang2019bertscore} together with BLEU-2 and BLEU-4~\cite{papineni2002bleu} to measure content preservation; this combination follows the evaluation protocol adopted in REMARK-LLM~\cite{zhang2024remark}. In the audio space, we reported Fr\'echet Audio Distance (FAD)~\cite{roblek2019fr}, following the fidelity evaluation used in ALIGNED-IS~\cite{wurobust}, to assess distribution-level audio quality. In addition, we further included speaker similarity, computed as the cosine similarity between speaker embeddings extracted by WavLM-Base-SV~\cite{chen2022wavlm}, as an extra metric to explicitly measure speaker consistency between the original and watermarked speeches.

\paragraph{Robustness evaluation.}
To evaluate robustness, we considered a post-processing threat model in which a watermarked speech sample may undergo common signal-processing or codec-based transformations before detection. Following AudioMarkBench~\cite{liu2024audiomarkbench}, we evaluated the \emph{no-box perturbation} categories. For signal-processing perturbations, we considered Gaussian noise, low-pass filtering, high-pass filtering, speed perturbation, smoothing, and echo. For compression and neural codec perturbations, we considered MP3, quantization, SoundStream~\cite{zeghidour2021soundstream}, Opus~\cite{valin2016high}, and EnCodec~\cite{defossezhigh}. Detailed descriptions of all perturbation settings are provided in Appendix~\ref{sec:perturbation_details}. For each attack, we evaluated multiple distortion strengths, including settings beyond the default configurations, while preserving intelligible speech content. This setup enables a systematic evaluation of watermark robustness from mild to relatively strong perturbations. We reported both per-attack performance across strength levels and results averaged over the full attack suite.

\begin{table}[h]
\centering
\caption{Evaluation of speech fidelity. For SSTMark, results are reported under the fixed text watermark configuration $\tau_{\mathrm{word}}=0.7$. Speaker similarity and Fr\'echet Audio Distance (FAD) are used to evaluate speaker consistency and audio quality, respectively. $\uparrow$ indicates that a higher value is better, and $\downarrow$ indicates that a lower value is better.}
\label{tab:tts_fidelity}
\scalebox{0.9}{%
\begin{tabular}{lcc}
\toprule
\textbf{Configuration} & \textbf{Speaker Sim $\uparrow$} & \textbf{FAD $\downarrow$} \\
\midrule
AudiowMark~\cite{westerfeld2020audiowmark} & \textbf{0.999} & 0.027 \\
WavMark~\cite{chen2023wavmark} & 0.998 & 1.934 \\
Timbre~\cite{liu2024detecting} & 0.997 & 1.578 \\
AudioSeal~\cite{san2024proactive} & \textbf{0.999} & 0.067 \\
SSTMark (Ours) & 0.985 & \textbf{0.001} \\
\bottomrule
\end{tabular}%
}
\end{table}

\begin{table*}[t]
\centering
\caption{Effect of the word-level similarity threshold $\tau_{\mathrm{word}}$ on text preservation and resynthesized speech fidelity in the SSTMark pipeline. Lower $\tau_{\mathrm{word}}$ relaxes the word-level similarity constraint and generally results in stronger watermark strength. $\uparrow$ indicates that a higher value is more desirable, and $\downarrow$ indicates that a lower value is more desirable.}
\label{tab:text_to_speech_fidelity}
\scalebox{0.9}{%
\begin{tabular}{cccccc}
\toprule
\textbf{$\tau_{\mathrm{word}}$} & \textbf{Speaker Similarity $\uparrow$} & \textbf{FAD $\downarrow$} & \textbf{BERTScore $\uparrow$} & \textbf{BLEU-2 $\uparrow$} & \textbf{BLEU-4 $\uparrow$} \\
\midrule
0.9 & 0.985 & 0.002 & 0.984 & 0.858 & 0.808  \\
0.8 & 0.985 & 0.003 & 0.974 & 0.784 & 0.693  \\
0.7 & 0.985 & 0.001 & 0.961 & 0.714 & 0.588  \\
0.6 & 0.984 & 0.002 & 0.955 & 0.691 & 0.554  \\
\bottomrule
\end{tabular}%
}
\end{table*}

\begin{table*}[h]
\centering
\small
\setlength{\tabcolsep}{3.0pt}
\caption{Watermark robustness under signal-processing edits, reported as TPR@FPR (\%). Detection thresholds are first calibrated on clean non-watermarked speech to satisfy target FPRs of 1\% and 0.1\%, and are then fixed for all evaluated conditions. Here, \textit{No-atk} denotes the original watermarked speech without being subjected to any attack. The table reports the corresponding true positive rates (TPRs) on watermarked speech under each attack condition. Higher values indicate better robustness. For each condition, the best result is highlighted in \textbf{bold} and the second-best result is \underline{underlined}.}
\label{tab:robust_detection_signalprocess}
\scalebox{0.9}{%
\begin{tabular}{llccccccccccccccc}
\toprule
\multirow{2}{*}{\textbf{Method}} 
& \textbf{No-atk}
& \multicolumn{3}{c}{\textbf{Gaussian Noise}}
& \multicolumn{3}{c}{\textbf{Low-pass}}
& \multicolumn{3}{c}{\textbf{High-pass}}
& \multicolumn{3}{c}{\textbf{Speed}}
& \multicolumn{1}{c}{\textbf{Smooth}}
& \multicolumn{1}{c}{\textbf{Echo}}
& \textbf{Avg.} \\
\cmidrule(lr){2-2}
\cmidrule(lr){3-5}
\cmidrule(lr){6-8}
\cmidrule(lr){9-11}
\cmidrule(lr){12-14}
\cmidrule(lr){15-15}
\cmidrule(lr){16-16}
\cmidrule(lr){17-17}
& 
& \textbf{10 dB} & \textbf{5 dB} & \textbf{0 dB}
& \textbf{1000} & \textbf{500} & \textbf{250}
& \textbf{1000} & \textbf{2000} & \textbf{3000}
& \textbf{$\times$0.75} & \textbf{$\times$1.25} & \textbf{$\times$1.5}
& \textbf{wid=6} & \textbf{0.5s} \\
\midrule
AudiowMark~\cite{westerfeld2020audiowmark} & 100.0 & 0.0 & 0.0 & 0.0 & \textbf{100.0} & \textbf{100.0} & \textbf{100.0} & \textbf{100.0} & \textbf{100.0} & \textbf{100.0} & 0.0 & 0.0 & 0.0 & \textbf{100.0} & \textbf{100.0} & 57.1 \\
\midrule
\textbf{TPR@FPR=1\%} \\
WavMark~\cite{chen2023wavmark} & 100.0 & 17.0 & 16.2 & \underline{15.4} & \underline{99.4} & 17.9 & 0.0 & \textbf{100.0} & \textbf{100.0} & \textbf{100.0} & \underline{99.9} & \underline{99.9} & \underline{98.5} & \textbf{100.0} & \textbf{100.0} & 68.9 \\
Timbre~\cite{liu2024detecting} & 100.0 & 72.6 & 22.9 & 4.1 & \textbf{100.0} & \textbf{100.0} & \textbf{98.7} & \textbf{100.0} & \textbf{100.0} & \textbf{100.0} & \textbf{100.0} & \textbf{100.0} & \textbf{100.0} & \textbf{100.0} & \textbf{100.0} & \underline{85.6} \\
AudioSeal~\cite{san2024proactive}  & 100.0 & \textbf{100.0} & \underline{58.6} & 0.0 & \textbf{100.0} & 83.4 & 52.2 & \textbf{100.0} & \textbf{100.0} & 1.6 & 98.3 & 99.3 & 95.6 & \textbf{100.0} & \textbf{100.0} & 78.2 \\
SSTMark (Ours) & 91.0 & \underline{91.6} & \textbf{90.0} & \textbf{88.3} & 89.3 & \underline{90.2} & \underline{89.3} & \underline{91.3} & \underline{89.5} & \underline{91.0} & 89.5 & 91.6 & 91.0 & \underline{90.5} & \underline{89.9} & \textbf{90.2} \\
\midrule
\textbf{TPR@FPR=0.1\%} \\
WavMark~\cite{chen2023wavmark} & 100.0 & 8.9 & 8.0 & \underline{8.0} & \underline{98.1} & 9.3 & 0.0 & \textbf{100.0} & \textbf{100.0} & \textbf{100.0} &  \underline{99.8} & \underline{99.8} & \underline{95.5} & \textbf{100.0} & \textbf{100.0} & 66.2 \\
Timbre~\cite{liu2024detecting} & 100.0 & 44.1 & 5.9 & 1.3 & \textbf{100.0} & \textbf{100.0} & \textbf{87.6} & \textbf{100.0} & \textbf{100.0} & \textbf{100.0} &  \textbf{100.0} & \textbf{100.0} & \textbf{100.0} & \textbf{100.0} & \textbf{100.0} & \underline{81.4} \\
AudioSeal~\cite{san2024proactive} & 100.0 & \textbf{100.0} & \underline{38.0} & 0.0 & \textbf{100.0} & 51.8 & 31.6 & \textbf{100.0} & \textbf{100.0} & 0.2 & 93.6 & 94.7 & 84.6 & \textbf{100.0} & \underline{99.9} & 71.5 \\
SSTMark (Ours) & 84.7 & \underline{83.8} & \textbf{83.0} & \textbf{76.8} & 84.1 & \underline{85.6} & \underline{84.6} & \underline{85.2} & \underline{84.4} & \underline{85.0} & 82.3 & 84.2 & 82.3 & \underline{84.4} & 81.0 & \textbf{83.4} \\
\bottomrule
\end{tabular}%
}
\end{table*}

\begin{table*}[h]
\centering
\small
\setlength{\tabcolsep}{3.5pt}
\caption{Watermark robustness under compression and neural codec edits, reported as TPR@FPR (\%). Detection thresholds are first calibrated on clean non-watermarked speech to satisfy target FPRs of 1\% and 0.1\%, and are then fixed for all evaluated conditions. Here, \textit{No-atk} denotes the original watermarked speech without being subjected to any attack. The table reports the corresponding true positive rates (TPRs) on watermarked speech under each attack condition. Higher values indicate better robustness. For each condition, the best result is highlighted in \textbf{bold} and the second-best result is \underline{underlined}.}
\label{tab:robust_detection_codec}
\scalebox{0.9}{%
\begin{tabular}{llccccccccccc}
\toprule
\multirow{2}{*}{\textbf{Method}}
& \textbf{No-atk}
& \multicolumn{2}{c}{\textbf{MP3}}
& \multicolumn{2}{c}{\textbf{Quantization}}
& \multicolumn{2}{c}{\textbf{SoundStream}}
& \multicolumn{2}{c}{\textbf{Opus}}
& \multicolumn{1}{c}{\textbf{Encodec}}
& \textbf{Avg.} \\
\cmidrule(lr){2-2}
\cmidrule(lr){3-4}
\cmidrule(lr){5-6}
\cmidrule(lr){7-8}
\cmidrule(lr){9-10}
\cmidrule(lr){11-11}
\cmidrule(lr){12-12}
& 
& \textbf{32 kbps} & \textbf{16 kbps} & \textbf{16 bits}
& \textbf{4 bits} & \textbf{16 quant.} & \textbf{4 quant.}
& \textbf{48 kbps} & \textbf{16 kbps} & \textbf{24 kHz}
& \\
\midrule
AudiowMark~\cite{westerfeld2020audiowmark} & 100.0 & \textbf{100.0}  & 0.0 & \textbf{100.0} & 96.6 & 0.0 & 0.0 & \textbf{100.0} & \textbf{100.0} & 0.0 & 55.2 \\
\midrule
\textbf{TPR@FPR=1\%} \\
WavMark~\cite{chen2023wavmark} & 100.0 & \underline{99.7} & 0.0 & \textbf{100.0} & 17.5 & 0.7 & 1.3 & 2.5 & 1.9 & 8.0 & 25.7 \\
Timbre~\cite{liu2024detecting} & 100.0 & 99.3 & \textbf{100.0} & \underline{99.8} & 3.2 & 3.2 & 3.2 & \textbf{100.0} & \textbf{100.0}  & 52.1 & \underline{73.1} \\
AudioSeal~\cite{san2024proactive} & 100.0 & \textbf{100.0} & \underline{99.8} & \textbf{100.0} & \textbf{100.0} & \underline{31.8} & \underline{32.2} & 0.0 & 0.1 & \textbf{99.7} & 62.6 \\
SSTMark (Ours) & 91.0 & 89.6 & 88.9 & 90.2 & \underline{90.4} & \textbf{90.2} & \textbf{88.9} & \underline{90.6} & \underline{91.4} & \underline{90.2} & \textbf{90.0} \\
\midrule
\textbf{TPR@FPR=0.1\%} \\
WavMark~\cite{chen2023wavmark} & 100.0 & \underline{99.6} & 0.0 & \textbf{100.0} & 8.3 & 0.2 & 0.3 & 0.6 & 0.5 & 3.5 & 23.7 \\
Timbre~\cite{liu2024detecting} & 100.0 & 81.3 & \textbf{100.0} & \underline{97.7} & 0.7 & 0.7 & 0.7 & \textbf{100.0} & \textbf{100.0} & 18.4 & \underline{66.5} \\
AudioSeal~\cite{san2024proactive} & 100.0 & \textbf{100.0} & \underline{90.2} & \textbf{100.0} & \textbf{100.0} & \underline{9.3} & \underline{8.7} & 0.0 & 0.0 & \textbf{97.0} & 56.1 \\
SSTMark (Ours) & 84.7 & 84.0 & 80.5 & 83.8 & \underline{81.4} & \textbf{81.3} & \textbf{79.7} & \underline{84.4} & \underline{85.2} & \underline{82.8} & \textbf{82.6} \\
\bottomrule
\end{tabular}%
}
\vspace{-10pt}
\end{table*}

\subsection{Fidelity Evaluation Results}
We evaluated SSTMark from two complementary perspectives: speech fidelity and text preservation. Specifically, we examined (1) whether the watermarked speech remains close to the speech originally generated by the model while preserving speaker consistency, and (2) whether the watermarked transcript still preserves the original semantic content. This evaluation is particularly important for SSTMark, as it embeds watermark through controlled lexical substitution rather than direct signal-level perturbation.

Table~\ref{tab:tts_fidelity} compares the speech fidelity of SSTMark against existing watermarking baselines using Fr\'echet Audio Distance (FAD) and speaker similarity. Following Aligned-IS~\cite{wurobust}, we used FAD to assess overall audio quality, and additionally include speaker similarity to evaluate whether speaker identity is preserved after watermark injection and speech resynthesis. Under the fixed setting $\tau_{\mathrm{word}}=0.7$, SSTMark achieves the lowest FAD of 0.001, indicating excellent overall audio quality of the resynthesized watermarked speech. In terms of speaker consistency, SSTMark obtains a speaker similarity of 0.985. Although this value is slightly lower than those of signal-level baselines, which are all near 1.0, it remains high in absolute terms, indicating that watermark injection and speech resynthesis do not cause a substantial loss of speaker consistency.

To further analyze the effect of watermark strength on fidelity, Table~\ref{tab:text_to_speech_fidelity} shows the performance of SSTMark under different values of $\tau_{\mathrm{word}}$. Several trends can be observed. First, speaker similarity remains stable at 0.985 across all settings, suggesting that speaker characteristics are consistently preserved despite changes in watermark strength. Second, FAD remains extremely low across all settings, ranging only from 0.001 to 0.003, indicating that varying the watermark strength has little impact on the overall audio quality of the resynthesized speech. 
Third, the BLEU score reflects the relationship between watermark strength and transcript fidelity. As $\tau_{\mathrm{word}}$ decreases from 0.9 to 0.6, both BLEU-2 and BLEU-4 exhibit clear declines. This trend is consistent with the role of $\tau_{\mathrm{word}}$ in the watermark injection stage: allowing more flexible synonym substitutions tends to strengthen the watermark, but also introduces larger lexical deviations from the original transcript. Although the decrease in BLEU scores reflects greater deviation of the watermarked text from the original wording, this does not directly conflict with our objective. Our goal is not to preserve the original speech content verbatim, but to inject watermark information while maintaining semantic similarity. Based on these concerns, we used $\tau_{\mathrm{word}}=0.7$ as the default setting in the subsequent experiments. This choice provides a favorable operating point, yielding a strong watermark while still maintaining high semantic similarity and speaker consistency.

\subsection{Robustness Evaluation Results}

We evaluated the robustness of SSTMark under the no-box perturbation setting defined in AudioMarkBench. Specifically, we considered a diverse set of perturbations, including both signal-processing edits and compression-related edits, and examine whether the watermark remains verifiable under these conditions. Following the calibration protocol described in Sec.~\ref{sec:experimental_setup}, detection thresholds for all methods were first determined on clean non-watermarked speech to satisfy fixed false positive rates (FPRs) of 1\% and 0.1\%, and were then kept fixed for all attacked conditions. This evaluation protocol reflects a realistic deployment scenario in which the detector does not know in advance whether an incoming audio sample has been attacked or what type of distortion has been applied. Robustness was shown as the true positive rate (TPR) under each attack condition, where a higher TPR indicates stronger robustness.

Table~\ref{tab:robust_detection_signalprocess} shows the results under signal-processing edits, including Gaussian noise, low-pass filtering, high-pass filtering, speed perturbation, smoothing, and echo.
The proposed SSTMark achieves the highest average TPR among all compared learnable baselines, reaching 90.2\% at TPR@FPR=1\% and 83.4\% at TPR@FPR=0.1\%. Notably, under the challenging Gaussian noise attack at 5 dB or even 0 dB, where several baselines degrade substantially, SSTMark remains highly stable, showing only limited drops relative to the no-attack case. While SSTMark does not deliver the strongest robustness under every individual condition, and its no-attack TPR is lower than that of several baselines, it achieves the best overall robustness across diverse attack scenarios. A possible explanation is that our text watermarking scheme inherently requires sufficiently long text to provide enough linguistic space for watermark injection.
%
Overall, these results support our earlier motivation: even when the speech waveform is perturbed by signal-processing operations, the semantic content it conveys is often largely preserved, allowing watermark embedded at the semantic level to remain robust for reliable detection.  

Table~\ref{tab:robust_detection_codec} shows the robustness results under compression and neural codec. The merit of SSTMark becomes even more pronounced in this setting. Specifically, SSTMark also achieves the highest average TPR among all compared learnable baselines, reaching 90.0\% at TPR@FPR=1\% and 82.6\% at TPR@FPR=0.1\%, while remaining highly stable across codec-related conditions with only moderate variation across bitrates and quantization levels. Notably, the compression and neural codec perturbations introduced in AudioMarkBench cover a diverse range of practical transmission formats, such as Opus~\cite{valin2016high}. Even under such conditions, SSTMark maintains stable detection performance. Although Timbre achieves the best results under Opus compression, SSTMark remains competitive without a substantial drop. Under SoundStream, SSTMark further achieves the best performance among all compared learnable baselines. Overall, these results show that SSTMark remains robust even when the audio undergoes severe codec-induced distortions.

Taken together, the results in Tables~\ref{tab:robust_detection_signalprocess} and \ref{tab:robust_detection_codec} demonstrate that SSTMark achieves the strongest average robustness profile across different perturbation types and strengths in the benchmark. Its main benefit is not that it wins in every single condition, but that it maintains comparatively stable detection performance without catastrophic degradation across a wide variety of edits. This robustness profile is especially desirable in realistic deployment scenarios, where the type and severity of post-processing cannot be assumed in advance. We further evaluate the generality of SSTMark by replacing its STT and TTS modules with alternative models. The corresponding results are provided in Appendix~\ref{sec:ablation_study}.

\begin{table}[t]
\centering
\small
\setlength{\tabcolsep}{3pt}
\caption{Fidelity of successful adversarial examples generated by HopSkipJumpAttack (HSJA) under different query budgets in terms of PESQ, SI-SNR, and ViSQOL. 
Higher values indicate that the attack can suppress watermark detection with less perceptual distortion, and vice versa.}
\label{tab:hsja_fidelity}
\scalebox{0.9}{%
\begin{tabular}{lcccc}
\toprule
\textbf{Method} & \textbf{Query Budget} & \textbf{PESQ} & \textbf{SI-SNR} & \textbf{ViSQOL} \\
\midrule
\multirow{3}{*}{SSTMark (Ours)}
& 100 & 1.142 & 2.688 & 1.652 \\
& 300 & 1.252 & 4.619 & 1.797 \\
& 500 & 1.235 & 4.132 & 1.785 \\
\midrule
\multirow{3}{*}{AudioSeal}
& 100 & 4.088 & 48.420 & 4.696 \\
& 300 & 4.365 & 55.511 & 4.726 \\
& 500 & 4.416 & 59.405 & 4.756 \\
\bottomrule
\end{tabular}}%
\end{table}

\subsection{Adversarial Attack}

We further evaluated robustness under adversarial watermark-removal attacks. Following AudioMarkBench, we adopted HopSkipJumpAttack (HSJA)~\cite{chen2020hopskipjumpattack}. HSJA is a decision-based boundary attack that first finds an initial adversarial example that flips the detector decision, typically via random or Gaussian-noise-based initialization, and then iteratively refines this example to approach the detector decision boundary while minimizing its distance to the original audio. In our experiments, HSJA was conducted in the waveform domain. To assess speech fidelity after attack, we adopted PESQ, SI-SNR, and ViSQOL. We compared SSTMark and AudioSeal under query budgets of $Q=100$, $300$, and $500$, where the query budget refers to the maximum number of times the attacker is allowed to query the detector during the attacking process.

As shown in Table~\ref{tab:hsja_fidelity}, SSTMark consistently exhibits much lower PESQ, SI-SNR, and ViSQOL than AudioSeal after HSJA across all query budgets. 
Since these metrics are computed between the attacked audio and its corresponding watermarked audio, the lower scores indicate that the adversary must introduce substantially larger perturbations to successfully remove the watermark from SSTMark. For example, at $Q=100$, the attacked samples of SSTMark only retain an SI-SNR of 2.688, whereas AudioSeal remains at 48.420, showing that successful attacks on SSTMark require far greater signal-level degradation. A similar trend is observed at $Q=300$ and $500$. Overall, these results indicate that SSTMark is more resistant to HSJA, as watermark removal requires substantially larger perceptual and signal-level changes to the audio.

\section{Conclusion and Limitations}

In this paper, we have proposed SSTMark, a training-free semantic-level speech watermarking framework. SSTMark binds watermark evidence to the semantic content conveyed by generated speech through speech transcription, text watermarking, and speech resynthesis. Experimental results on AudioMarkBench show that SSTMark remains stable across different perturbation types and strengths without severe detection-rate collapse, suggesting that semantic-level watermarking is a promising direction for robust provenance verification of generated speech.

Despite the aforementioned advantages, a further study for SSTMark is to extend the current framework to shorter speech utterances and improve its robustness and detection performance in short-duration speech scenarios.


\clearpage
\newpage
\bibliographystyle{ACM-Reference-Format}
\bibliography{sample-base}

\clearpage
\newpage
\appendix

\section{Pseudocode for SSTMark}
\label{sec:pseudocode}

Algorithm~\ref{alg:sstmark_injection} summarizes the watermark injection pipeline of SSTMark. The keyed text watermarking operator $\mathbf{W}(\cdot;k,\tau_{\mathrm{sent}},\tau_{\mathrm{word}})$ in Eq.~\eqref{eq:watermark} follows the synonym-based rewriting procedure in Sec.~\ref{sec:prelim_text_wm}, with keyed binary encoding defined in Eq.~\eqref{eq:keyed_token_bit} and Eq.~\eqref{eq:keyed_synonym_bit}, and candidate filtering defined in Eq.~\eqref{eq:candidate_filter}.

Algorithm~\ref{alg:sstmark_detection} summarizes the watermark detection pipeline of SSTMark. The keyed text watermark detector $\mathbf{D}(\cdot;k)$ in Eq.~\eqref{eq:wm_detector} follows the hypothesis-testing procedure in Sec.~\ref{sec:prelim_text_wm}, where watermark evidence is evaluated under the keyed encoding in Eq.~\eqref{eq:keyed_token_bit} using the z-score formulation in Eq.~\eqref{eq:zscore}.

\begin{algorithm}[h]
\caption{SSTMark watermark injection}
\label{alg:sstmark_injection}
\begin{algorithmic}[1]
\Require Prompt $p$, speech generator $\mathbf{G}_{\mathrm{sp}}(\cdot)$, ASR model $\mathbf{A}(\cdot)$, keyed text watermarking operator $\mathbf{W}(\cdot;k,\tau_{\mathrm{sent}},\tau_{\mathrm{word}})$, TTS model $\mathbf{T}(\cdot)$, watermark key $k$, sentence-level threshold $\tau_{\mathrm{sent}}$, word-level threshold $\tau_{\mathrm{word}}$
\Ensure Watermarked speech $s_{\mathrm{wm}}$

\State $s_{\mathrm{pre}} \gets \mathbf{G}_{\mathrm{sp}}(p)$ \Comment{Eq.~\eqref{eq:pre-speech}}
\State $x_{\mathrm{pre}} \gets \mathbf{A}(s_{\mathrm{pre}})$ \Comment{Eq.~\eqref{eq:STT}}

\State Instantiate the keyed text watermarking operator under semantic constraints $\tau_{\mathrm{sent}}$ and $\tau_{\mathrm{word}}$
\State Configure $\mathbf{W}(\cdot;k,\tau_{\mathrm{sent}},\tau_{\mathrm{word}})$ with the keyed bit assignment rules in Eq.~\eqref{eq:keyed_token_bit} and Eq.~\eqref{eq:keyed_synonym_bit}, and the candidate filtering rule in Eq.~\eqref{eq:candidate_filter}
\State $x_{\mathrm{wm}} \gets \mathbf{W}(x_{\mathrm{pre}};k,\tau_{\mathrm{sent}},\tau_{\mathrm{word}})$ \Comment{Eq.~\eqref{eq:watermark}}

\State $s_{\mathrm{wm}} \gets \mathbf{T}(x_{\mathrm{wm}})$ \Comment{Eq.~\eqref{eq:TTS}}
\State \Return $s_{\mathrm{wm}}$
\end{algorithmic}
\end{algorithm}

\begin{algorithm}[h]
\caption{SSTMark watermark detection}
\label{alg:sstmark_detection}
\begin{algorithmic}[1]
\Require Suspect speech $s_{\mathrm{sus}}$, ASR model $\mathbf{A}(\cdot)$, keyed text watermark detector $\mathbf{D}(\cdot;k)$, watermark key $k$
\Ensure Watermark score $z$

\State $x_{\mathrm{sus}} \gets \mathbf{A}(s_{\mathrm{sus}})$ \Comment{Eq.~\eqref{eq:STT_sus}}
\State Instantiate the keyed text watermark detector with watermark key $k$
\State Configure $\mathbf{D}(\cdot;k)$ to compute keyed token encodings according to Eq.~\eqref{eq:keyed_token_bit} and quantify watermark evidence using Eq.~\eqref{eq:zscore}
\State $z \gets \mathbf{D}(x_{\mathrm{sus}};k)$ \Comment{Eq.~\eqref{eq:wm_detector}}
\State \Return $z$
\end{algorithmic}
\end{algorithm}

\section{Speech Generation Instruction Template}
\label{sec:generation_template}
For reproducibility, we provide below the instruction template used for SpeechGPT-based speech generation. Following the original SpeechGPT setup~\cite{zhang2023speechgpt}, each text prompt from the LibriSpeech train-clean-100 split is inserted into the placeholder \texttt{[TEXT]}.

\begin{quote}
\small
"You are an AI assistant whose name is SpeechGPT. SpeechGPT is an intrinsic cross-modal conversational language model developed by Fudan University. SpeechGPT can understand and communicate fluently with humans through speech or text chosen by the user. It can perceive cross-modal inputs and generate cross-modal outputs. [Human]: Read this sentence aloud. This is the input: \texttt{[TEXT]} <eoh>. [SpeechGPT]:"
\end{quote}

\subsection{Anonymous Audio Demo Samples}
We provide 10 anonymous demo audio files in the supplementary material. These samples consist of 5 paired examples, each containing two versions of audio: one non-watermarked sample and one watermarked sample generated by the proposed method. For each pair, \texttt{Sample\_XX\_A.wav} denotes the non-watermarked speech sample, while \texttt{Sample\_XX\_B.wav} denotes the corresponding watermarked speech sample generated by the proposed method.

All demo files are anonymized and named using neutral identifiers. The filenames, folder names, and accompanying descriptions do not contain author names, institution names, project names, repository links, or any other information that may reveal the identity of the authors. In addition, the audio files are distributed without identifying metadata.

\section{Details of Audio Perturbations}
\label{sec:perturbation_details}

Our perturbation setup is based on AudioMarkBench~\cite{liu2024audiomarkbench}. In addition to the benchmark settings, we further include a few stronger perturbation levels beyond the benchmark range to assess robustness in more aggressive scenarios. Below, we summarize the benchmark settings and our additional settings for each perturbation.

\begin{itemize}
    \item \textbf{Gaussian noise:} Additive Gaussian noise is applied to the waveform. Strength parameter: target SNR. Benchmark range: [5, 40] dB. Our experimental settings: 10, 5, and 0 dB.

    \item \textbf{Low-pass filter:} A low-pass filter removes high-frequency components. Strength parameter: cutoff frequency. Benchmark range: [800, 4000] Hz. Our experimental settings: 1000, 500, and 250 Hz.

    \item \textbf{High-pass filter:} A high-pass filter removes low-frequency components. Strength parameter: cutoff frequency. Benchmark range: [800, 4000] Hz. Our experimental settings: 1000, 2000, and 3000 Hz.

    \item \textbf{Speed perturbation:} Playback speed is modified. Strength parameter: speed factor. Benchmark range: [0.7, 1.5]. Our experimental settings: $\times$0.75, $\times$1.25, and $\times$1.5.

    \item \textbf{Smoothing:} Temporal smoothing suppresses rapid local waveform variations. Strength parameter: window size. Benchmark range: [6, 22]. Our experimental setting: width $=6$.

    \item \textbf{Echo:} Echo adds a delayed and attenuated copy of the waveform. Strength parameter: delay. Benchmark range: [0.1, 0.9] s. Our experimental setting: 0.5 s, with decay factor fixed at 0.5.

    \item \textbf{MP3:} MP3 applies lossy audio compression. Strength parameter: bitrate. Benchmark range: [8, 40] kbps. Our experimental settings: 32 and 16 kbps.

    \item \textbf{Quantization:} Quantization reduces waveform precision. Strength parameter: bit depth. Benchmark range: [4, 64]. Our experimental settings: 16-bit and 4-bit quantization.

    \item \textbf{SoundStream:} SoundStream~\cite{zeghidour2021soundstream} applies neural audio compression. Strength parameter: number of quantizers. Benchmark range: [4, 16]. Our experimental settings: 16 and 4 quantizers.

    \item \textbf{Opus:} Opus~\cite{valin2016high} applies perceptual audio compression. Strength parameter: bitrate. Benchmark range: [16, 256] kbps. Our experimental settings: 48 and 16 kbps.

    \item \textbf{EnCodec:} EnCodec~\cite{defossezhigh} applies neural audio compression. Strength parameter: bandwidth. Benchmark range: [1.5, 24.0] kHz. Our experimental setting: 24 kHz.
\end{itemize}

\section{Ablation Study}
\label{sec:ablation_study}

\begin{table*}[h]
\centering
\small
\setlength{\tabcolsep}{3.0pt}
\caption{Ablation study of the TTS backbone in SSTMark: Watermark robustness under signal-processing edits, reported as TPR@FPR (\%). For SSTMark, the original TTS model CosyVoice~\cite{du2024cosyvoice} is replaced with F5-TTS~\cite{chen2025f5}. Detection thresholds are first calibrated on clean non-watermarked speech to satisfy target FPRs of 1\% and 0.1\%, and are then fixed for all evaluated conditions. Here, \textit{No-atk} denotes the original watermarked speech without being subjected to any attack. The table reports the corresponding true positive rates (TPRs) on watermarked speech under each attack condition. Higher values indicate better robustness. For each condition, the best result is highlighted in \textbf{bold} and the second-best result is \underline{underlined}.}
\label{tab:f5-tts_signalprocess}
\scalebox{0.9}{%
\begin{tabular}{llccccccccccccccc}
\toprule
\multirow{2}{*}{\textbf{Method}} 
& \textbf{No-atk}
& \multicolumn{3}{c}{\textbf{Gaussian Noise}}
& \multicolumn{3}{c}{\textbf{Low-pass}}
& \multicolumn{3}{c}{\textbf{High-pass}}
& \multicolumn{3}{c}{\textbf{Speed}}
& \multicolumn{1}{c}{\textbf{Smooth}}
& \multicolumn{1}{c}{\textbf{Echo}}
& \textbf{Avg.} \\
\cmidrule(lr){2-2}
\cmidrule(lr){3-5}
\cmidrule(lr){6-8}
\cmidrule(lr){9-11}
\cmidrule(lr){12-14}
\cmidrule(lr){15-15}
\cmidrule(lr){16-16}
\cmidrule(lr){17-17}
& 
& \textbf{10 dB} & \textbf{5 dB} & \textbf{0 dB}
& \textbf{1000} & \textbf{500} & \textbf{250}
& \textbf{1000} & \textbf{2000} & \textbf{3000}
& \textbf{$\times$0.75} & \textbf{$\times$1.25} & \textbf{$\times$1.5}
& \textbf{wid=6} & \textbf{0.5s} \\
\midrule
AudiowMark~\cite{westerfeld2020audiowmark} & 100.0 & 0.0 & 0.0 & 0.0 & 100.0 & 100.0 & 100.0 & 100.0 & 100.0 & 100.0 & 0.0 & 0.0 & 0.0 & 100.0 & 100.0 & 57.1 \\
\midrule
\textbf{TPR@FPR=1\%} \\
WavMark~\cite{chen2023wavmark} & 100.0 & 17.0 & 16.2 & \underline{15.4} & \underline{99.4} & 17.9 & 0.0 & \textbf{100.0} & \textbf{100.0} & \textbf{100.0} & \underline{99.9} & \underline{99.9} & \underline{98.5} & \textbf{100.0} & \textbf{100.0} & 68.9 \\
Timbre~\cite{liu2024detecting} & 100.0 & 72.6 & 22.9 & 4.1 & \textbf{100.0} & \textbf{100.0} & \textbf{98.7} & \textbf{100.0} & \textbf{100.0} & \textbf{100.0} & \textbf{100.0} & \textbf{100.0} & \textbf{100.0} & \textbf{100.0} & \textbf{100.0} & \textbf{85.6} \\
AudioSeal~\cite{san2024proactive}  & 100.0 & \textbf{100.0} & \underline{58.6} & 0.0 & \textbf{100.0} & 83.4 & 52.2 & \textbf{100.0} & \textbf{100.0} & 1.6 & 98.3 & 99.3 & 95.6 & \textbf{100.0} & \textbf{100.0} & 78.2 \\
SSTMark (Ours) & 85.9 & \underline{86.5} & \textbf{84.2} & \textbf{80.2} & 84.0 & \underline{84.0} & \underline{96.1} & \underline{85.6} & \underline{86.7} & \underline{87.3} & 84.8 & 83.7 & 84.6 & \underline{86.8} & \underline{81.4} & \underline{84.7} \\
\midrule
\textbf{TPR@FPR=0.1\%} \\
WavMark~\cite{chen2023wavmark} & 100.0 & 8.9 & 8.0 & \underline{8.0} & \underline{98.1} & 9.3 & 0.0 & \textbf{100.0} & \textbf{100.0} & \textbf{100.0} & \underline{99.8} & \underline{99.8} & \underline{95.5} & \textbf{100.0} & \textbf{100.0} & 66.2 \\
Timbre~\cite{liu2024detecting} & 100.0 & 44.1 & 5.9 & 1.3 & \textbf{100.0} & \textbf{100.0} & \textbf{87.6} & \textbf{100.0} & \textbf{100.0} & \textbf{100.0} & \textbf{100.0} & \textbf{100.0} & \textbf{100.0} & \textbf{100.0} & \textbf{100.0} & \textbf{81.4} \\
AudioSeal~\cite{san2024proactive} & 100.0 & \textbf{100.0} & \underline{38.0} & 0.0 & \textbf{100.0} & 51.8 & 31.6 & \textbf{100.0} & \textbf{100.0} & 0.2 & 93.6 & 94.7 & 84.6 & \textbf{100.0} & \underline{99.9} & 71.5 \\
SSTMark (Ours) & 77.1 & \underline{73.8} & \textbf{73.2} & \textbf{68.4} & 75.8 & \underline{75.3} & \underline{78.8} & \underline{78.3} & \underline{79.0} & \underline{79.3} & 76.4 & 74.9 & 75.5 & \underline{79.0} & 68.2 & \underline{75.4} \\
\bottomrule
\end{tabular}%
}
\end{table*}

\begin{table*}[h]
\centering
\small
\setlength{\tabcolsep}{3.5pt}
\caption{Ablation study of the TTS backbone in SSTMark: Watermark robustness under compression and neural codec edits, reported as TPR@FPR (\%). For SSTMark, the original TTS model CosyVoice~\cite{du2024cosyvoice} is replaced with F5-TTS~\cite{chen2025f5}. Detection thresholds are first calibrated on clean non-watermarked speech to satisfy target FPRs of 1\% and 0.1\%, and are then fixed for all evaluated conditions. Here, \textit{No-atk} denotes the original watermarked speech without being subjected to any attack. The table reports the corresponding true positive rates (TPRs) on watermarked speech under each attack condition. Higher values indicate better robustness. For each condition, the best result is highlighted in \textbf{bold} and the second-best result is \underline{underlined}.}
\label{tab:f5-tts_codec}
\scalebox{0.9}{%
\begin{tabular}{llccccccccccc}
\toprule
\multirow{2}{*}{\textbf{Method}}
& \textbf{No-atk}
& \multicolumn{2}{c}{\textbf{MP3}}
& \multicolumn{2}{c}{\textbf{Quantization}}
& \multicolumn{2}{c}{\textbf{SoundStream}}
& \multicolumn{2}{c}{\textbf{Opus}}
& \multicolumn{1}{c}{\textbf{Encodec}}
& \textbf{Avg.} \\
\cmidrule(lr){2-2}
\cmidrule(lr){3-4}
\cmidrule(lr){5-6}
\cmidrule(lr){7-8}
\cmidrule(lr){9-10}
\cmidrule(lr){11-11}
\cmidrule(lr){12-12}
& 
& \textbf{32 kbps} & \textbf{16 kbps} & \textbf{16 bits}
& \textbf{4 bits} & \textbf{16 quant.} & \textbf{4 quant.}
& \textbf{48 kbps} & \textbf{16 kbps} & \textbf{24 kHz}
& \\
\midrule
AudiowMark~\cite{westerfeld2020audiowmark} & 100.0 & 100.0  & 0.0 & 100.0 & 96.6 & 0.0 & 0.0 & 100.0 & 100.0 & 0.0 & 55.2 \\
\midrule
\textbf{TPR@FPR=1\%} \\
WavMark~\cite{chen2023wavmark} & 100.0 & \underline{99.7} & 0.0 & \textbf{100.0} & 17.5 & 0.7 & 1.3 & 2.5 & 1.9 & 8.0 & 25.7 \\
Timbre~\cite{liu2024detecting} & 100.0 & 99.3 & \textbf{100.0} & \underline{99.8} & 3.2 & 3.2 & 3.2 & \textbf{100.0} & \textbf{100.0}  & 52.1 & \underline{73.1} \\
AudioSeal~\cite{san2024proactive} & 100.0 & \textbf{100.0} & \underline{99.8} & \textbf{100.0} & \textbf{100.0} & \underline{31.8} & \underline{32.2} & 0.0 & 0.1 & \textbf{99.7} & 62.6 \\
SSTMark (Ours) & 85.9 & 86.0 & 80.6 & 84.6 & \underline{82.6} & \textbf{81.0} & \textbf{81.6} & \underline{86.2} & \underline{85.4} & \underline{82.5} & \textbf{83.4} \\
\midrule
\textbf{TPR@FPR=0.1\%} \\
WavMark~\cite{chen2023wavmark} & 100.0 & \underline{99.6} & 0.0 & \textbf{100.0} & 8.3 & 0.2 & 0.3 & 0.6 & 0.5 & 3.5 & 23.7 \\
Timbre~\cite{liu2024detecting} & 100.0 & 81.3 & \textbf{100.0} & \underline{97.7} & 0.7 & 0.7 & 0.7 & \textbf{100.0} & \textbf{100.0} & 18.4 & \underline{66.5} \\
AudioSeal~\cite{san2024proactive} & 100.0 & \textbf{100.0} & \underline{90.2} & \textbf{100.0} & \textbf{100.0} & \underline{9.3} & \underline{8.7} & 0.0 & 0.0 & \textbf{97.0} & 56.1 \\
SSTMark (Ours) & 77.1 & 76.2 & 69.0 & 78.9 & \underline{69.1} & \textbf{70.2} & \textbf{72.6} & \underline{79.8} & \underline{76.4} & \underline{71.7} & \textbf{73.8} \\
\bottomrule
\end{tabular}%
}
\end{table*}

\begin{table*}[h]
\centering
\small
\setlength{\tabcolsep}{3.0pt}
\caption{Ablation study of the STT backbone in SSTMark: watermark robustness under signal-processing edits, reported as TPR@FPR (\%). For SSTMark, the original STT model Whisper is replaced with Canary-1B-v2. Detection thresholds are first calibrated on clean non-watermarked speech to satisfy target FPRs of 1\% and 0.1\%, and are then fixed for all evaluated conditions. Here, \textit{No-atk} denotes the original watermarked speech without being subjected to any attack. The table reports the corresponding true positive rates (TPRs) on watermarked speech under each attack condition. Higher values indicate better robustness. For each condition, the best result is highlighted in \textbf{bold} and the second-best result is \underline{underlined}.}
\label{tab:canary-stt_signalprocess}
\scalebox{0.9}{%
\begin{tabular}{llccccccccccccccc}
\toprule
\multirow{2}{*}{\textbf{Method}} 
& \textbf{No-atk}
& \multicolumn{3}{c}{\textbf{Gaussian Noise}}
& \multicolumn{3}{c}{\textbf{Low-pass}}
& \multicolumn{3}{c}{\textbf{High-pass}}
& \multicolumn{3}{c}{\textbf{Speed}}
& \multicolumn{1}{c}{\textbf{Smooth}}
& \multicolumn{1}{c}{\textbf{Echo}}
& \textbf{Avg.} \\
\cmidrule(lr){2-2}
\cmidrule(lr){3-5}
\cmidrule(lr){6-8}
\cmidrule(lr){9-11}
\cmidrule(lr){12-14}
\cmidrule(lr){15-15}
\cmidrule(lr){16-16}
\cmidrule(lr){17-17}
& 
& \textbf{10 dB} & \textbf{5 dB} & \textbf{0 dB}
& \textbf{1000} & \textbf{500} & \textbf{250}
& \textbf{1000} & \textbf{2000} & \textbf{3000}
& \textbf{$\times$0.75} & \textbf{$\times$1.25} & \textbf{$\times$1.5}
& \textbf{wid=6} & \textbf{0.5s} \\
\midrule
AudiowMark~\cite{westerfeld2020audiowmark} & 100.0 & 0.0 & 0.0 & 0.0 & 100.0 & 100.0 & 100.0 & 100.0 & 100.0 & 100.0 & 0.0 & 0.0 & 0.0 & 100.0 & 100.0 & 57.1 \\
\midrule
\textbf{TPR@FPR=1\%} \\
WavMark~\cite{chen2023wavmark} & 100.0 & 17.0 & 16.2 & \underline{15.4} & \underline{99.4} & 17.9 & 0.0 & \textbf{100.0} & \textbf{100.0} & \textbf{100.0} & \underline{99.9} & \underline{99.9} & \underline{98.5} & \textbf{100.0} & \textbf{100.0} & 68.9 \\
Timbre~\cite{liu2024detecting} & 100.0 & 72.6 & 22.9 & 4.1 & \textbf{100.0} & \textbf{100.0} & \textbf{98.7} & \textbf{100.0} & \textbf{100.0} & \textbf{100.0} & \textbf{100.0} & \textbf{100.0} & \textbf{100.0} & \textbf{100.0} & \textbf{100.0} & \textbf{85.6} \\
AudioSeal~\cite{san2024proactive}  & 100.0 & \textbf{100.0} & \underline{58.6} & 0.0 & \textbf{100.0} & 83.4 & 52.2 & \textbf{100.0} & \textbf{100.0} & 1.6 & 98.3 & 99.3 & 95.6 & \textbf{100.0} & \textbf{100.0} & 78.2 \\
SSTMark (Ours) & 86.0 & \underline{88.7} & \textbf{87.3} & \textbf{83.5} & 86.5 & \underline{86.1} & \underline{86.5} & \underline{85.8} & \underline{85.1} & \underline{84.6} & 83.7 & 84.0 & 84.2 & \underline{85.5} & \underline{74.3} & \underline{84.7} \\
\midrule
\textbf{TPR@FPR=0.1\%} \\
WavMark~\cite{chen2023wavmark} & 100.0 & 8.9 & 8.0 & \underline{8.0} & \underline{98.1} & 9.3 & 0.0 & \textbf{100.0} & \textbf{100.0} & \textbf{100.0} & \underline{99.8} & \underline{99.8} & \underline{95.5} & \textbf{100.0} & \textbf{100.0} & 66.2 \\
Timbre~\cite{liu2024detecting} & 100.0 & 44.1 & 5.9 & 1.3 & \textbf{100.0} & \textbf{100.0} & \textbf{87.6} & \textbf{100.0} & \textbf{100.0} & \textbf{100.0} & \textbf{100.0} & \textbf{100.0} & \textbf{100.0} & \textbf{100.0} & \textbf{100.0} & \textbf{81.4} \\
AudioSeal~\cite{san2024proactive} & 100.0 & \textbf{100.0} & \underline{38.0} & 0.0 & \textbf{100.0} & 51.8 & 31.6 & \textbf{100.0} & \textbf{100.0} & 0.2 & 93.6 & 94.7 & 84.6 & \textbf{100.0} & \underline{99.9} & 71.5 \\
SSTMark (Ours) & 75.6 & \underline{78.5} & \textbf{74.6} & \textbf{67.8} & 77.0 & \underline{75.9} & \underline{75.7} & \underline{74.8} & \underline{74.3} & \underline{73.6} & 72.8 & 72.1 & 73.2 & \underline{76.2} & 55.4 & \underline{73.0} \\
\bottomrule
\end{tabular}%
}
\end{table*}

\begin{table*}[h]
\centering
\small
\setlength{\tabcolsep}{3.5pt}
\caption{Ablation study of the STT backbone in SSTMark: watermark robustness under compression and neural codec edits, reported as TPR@FPR (\%). For SSTMark, the original STT model Whisper is replaced with Canary-1B-v2. Detection thresholds are first calibrated on clean non-watermarked speech to satisfy target FPRs of 1\% and 0.1\%, and are then fixed for all evaluated conditions. Here, \textit{No-atk} denotes the original watermarked speech without being subjected to any attack. The table reports the corresponding true positive rates (TPRs) on watermarked speech under each attack condition. Higher values indicate better robustness. For each condition, the best result is highlighted in \textbf{bold} and the second-best result is \underline{underlined}.}
\label{tab:canary-stt_codec}
\scalebox{0.9}{%
\begin{tabular}{llccccccccccc}
\toprule
\multirow{2}{*}{\textbf{Method}}
& \textbf{No-atk}
& \multicolumn{2}{c}{\textbf{MP3}}
& \multicolumn{2}{c}{\textbf{Quantization}}
& \multicolumn{2}{c}{\textbf{SoundStream}}
& \multicolumn{2}{c}{\textbf{Opus}}
& \multicolumn{1}{c}{\textbf{Encodec}}
& \textbf{Avg.} \\
\cmidrule(lr){2-2}
\cmidrule(lr){3-4}
\cmidrule(lr){5-6}
\cmidrule(lr){7-8}
\cmidrule(lr){9-10}
\cmidrule(lr){11-11}
\cmidrule(lr){12-12}
& 
& \textbf{32 kbps} & \textbf{16 kbps} & \textbf{16 bits}
& \textbf{4 bits} & \textbf{16 quant.} & \textbf{4 quant.}
& \textbf{48 kbps} & \textbf{16 kbps} & \textbf{24 kHz}
& \\
\midrule
AudiowMark~\cite{westerfeld2020audiowmark} & 100.0 & 100.0  & 0.0 & 100.0 & 96.6 & 0.0 & 0.0 & 100.0 & 100.0 & 0.0 & 55.2 \\
\midrule
\textbf{TPR@FPR=1\%} \\
WavMark~\cite{chen2023wavmark} & 100.0 & \underline{99.7} & 0.0 & \textbf{100.0} & 17.5 & 0.7 & 1.3 & 2.5 & 1.9 & 8.0 & 25.7 \\
Timbre~\cite{liu2024detecting} & 100.0 & 99.3 & \textbf{100.0} & \underline{99.8} & 3.2 & 3.2 & 3.2 & \textbf{100.0} & \textbf{100.0}  & \underline{52.1} & \underline{73.1} \\
AudioSeal~\cite{san2024proactive} & 100.0 & \textbf{100.0} & \underline{99.8} & \textbf{100.0} & \textbf{100.0} & \underline{31.8} & \underline{32.2} & 0.0 & 0.1 & \textbf{99.7} & 62.6 \\
SSTMark (Ours) & 86.0 & 84.5 & 85.1 & 84.9 & \underline{83.5} & \textbf{84.4} & \textbf{84.3} & \underline{85.2} & \underline{83.4} & 83.9 & \textbf{84.4} \\
\midrule
\textbf{TPR@FPR=0.1\%} \\
WavMark~\cite{chen2023wavmark} & 100.0 & \underline{99.6} & 0.0 & \textbf{100.0} & 8.3 & 0.2 & 0.3 & 0.6 & 0.5 & 3.5 & 23.7 \\
Timbre~\cite{liu2024detecting} & 100.0 & 81.3 & \textbf{100.0} & \underline{97.7} & 0.7 & 0.7 & 0.7 & \textbf{100.0} & \textbf{100.0} & \underline{18.4} & \underline{66.5} \\
AudioSeal~\cite{san2024proactive} & 100.0 & \textbf{100.0} & \underline{90.2} & \textbf{100.0} & \textbf{100.0} & \underline{9.3} & \underline{8.7} & 0.0 & 0.0 & \textbf{97.0} & 56.1 \\
SSTMark (Ours) & 75.6 & 74.4 & 73.6 & 74.2 & \underline{72.2} & \textbf{74.8} & \textbf{74.8} & \underline{74.9} & \underline{73.2} & 73.6 & \textbf{74.0} \\
\bottomrule
\end{tabular}%
}
\end{table*}

We investigate the robustness of SSTMark under backbone model replacement of both the TTS and STT modules. Specifically, we replace CosyVoice~\cite{du2024cosyvoice} with F5-TTS~\cite{chen2025f5}, and Whisper~\cite{radford2023robust} with Canary-1B-v2\footnote{Monica Sekoyan, Nithin Rao Koluguri, Nune Tadevosyan, Piotr Zelasko, Travis Bartley, Nikolay Karpov, Jagadeesh Balam, and Boris Ginsburg. "Canary-1B-v2 \& Parakeet-TDT-0.6B-v3: Efficient and High-Performance Models for Multilingual ASR and AST." \textit{arXiv preprint} arXiv:2509.14128, 2025.}.

\paragraph{TTS model replacement.}
Replacing CosyVoice~\cite{du2024cosyvoice} with F5-TTS~\cite{chen2025f5} still yields consistent robustness under both signal-processing and compression distortions at two operating points, as shown in Tables~\ref{tab:f5-tts_signalprocess} and~\ref{tab:f5-tts_codec}, respectively. In this setting, SSTMark achieves average TPRs of $84.7\%/75.4\%$ under signal-processing attacks and $83.4\%/73.8\%$ under codec-related attacks at FPRs of 1\%/0.1\%, respectively.

\paragraph{STT model replacement.}
Replacing the original Whisper STT module~\cite{radford2023robust} with Canary-1B-v2~\footnotemark[\value{footnote}] likewise preserves robust performance under diverse distortions. Specifically, SSTMark achieves average TPRs of $84.7\%/73.0\%$ under signal-processing edits and $84.4\%/74.0\%$ under codec-related edits at FPRs of 1\%/0.1\%, respectively.

\end{document}